\documentclass[aps,superscriptaddress,showpacs,reprint]{revtex4-1}
\usepackage{color}
\usepackage{natbib}
\usepackage{amsmath}
\usepackage{amssymb}
\usepackage[pdftex]{graphicx}
\usepackage{epstopdf}
\usepackage[latin1]{inputenc}
\usepackage[ngerman, english]{babel} 
\newcommand{\EB}{$E_{\rm B}$}

\newcommand{\SB}{SmB$_{\rm 6}$}
\newcommand{\nf}{$n_{\rm f}$}
\newcommand{\vl}{$\textit{V}_{\rm Sm}$}
\newcommand{\stw}{Sm$^{\rm 2+}$}
\newcommand{\sth}{Sm$^{\rm 3+}$}

\begin{document}

%\jvol{00} \jnum{00} \jyear{2015} \jmonth{January}

%\articletype{GUIDE}

\title{Valence Characterization of Surface and Subsurface Region in SmB$_6$}
\date{\today}
\author{P. Lutz}
\affiliation{Experimentelle Physik VII and R\"ontgen Research Center for Complex Materials (RCCM), Universit\"at W\"urzburg,  97074 W\"urzburg, Germany}

\author{M. Thees}
\affiliation{Experimentelle Physik VII and R\"ontgen Research Center for Complex Materials (RCCM), Universit\"at W\"urzburg,  97074 W\"urzburg, Germany}

\author{T. R. F. Peixoto}
\affiliation{Experimentelle Physik VII and R\"ontgen Research Center for Complex Materials (RCCM), Universit\"at W\"urzburg,  97074 W\"urzburg, Germany}

\author{B. Y. Kang}
\affiliation{School of Materials Science and Engineering, Gwangju Institute of Science and Technology (GIST), Gwangju 61005, Korea.}

\author{B. K. Cho}
\affiliation{School of Materials Science and Engineering, Gwangju Institute of Science and Technology (GIST), Gwangju 61005, Korea.}

\author{Chul-Hee Min}
\email[corresponding author. Email:]{cmin@physik.uni-wuerzburg.de}
\affiliation{Experimentelle Physik VII and R\"ontgen Research Center for Complex Materials (RCCM), Universit\"at W\"urzburg,  97074 W\"urzburg, Germany}

\author{F. Reinert}
\affiliation{Experimentelle Physik VII and R\"ontgen Research Center for Complex Materials (RCCM), Universit\"at W\"urzburg,  97074 W\"urzburg, Germany}

\begin{abstract}
Samarium hexaboride (\SB), which lies in the mixed valence regime in the Anderson model, has been predicted to possess topologically protected surface states. The intensive investigations on \SB~have brought up the long standing questions about the discrepancy between the surface and bulk electronic properties in rare-earth compounds in general. Here, we investigate and eventually clarify this discrepancy in the particular case of \SB~by the photoemission core-level spectra. We focus on the change in both Sm and B states depending on time, temperature, probing depth, and surface termination on the cleaved (100) surface. Our spectra show that the unusual time-dependent change in the Sm valence occurs within a period of hours, which is not related to the adsorption of residual gases. Moreover, we observe a reduction of the surface feature in the B and Sm states on the same timescale accompanied by the formation of a subsurface region. Thus, it indicates the relatively slow charge redistribution between the surface and subsurface regions. Our findings demonstrate that the $f$\,states is strongly involved in the surface relaxation.
\end{abstract}

\maketitle
\section{Introduction}
\subsection{Charge fluctuation and topology}
Charge (valence) fluctuating many-body systems have gained a particular interest recently since samarium hexaboride (\SB), the first candidate for a topological Kondo insulator \cite{dzero_topological_2010}, belongs to the mixed valence regime of the Anderson model \cite{hewson__1993,werner_interaction_2013, martin_theory_1979,varma_mixed-valence_1976,min_importance_2014}. The unique properties of the insulating phase in this regime have been investigated such as the emergence of  the coherent band inversion, the gap opening, and in-gap states \cite{legner_topolgical_2014,werner_dynamically_2014,chen_optical_2014,min_universal_2015}. Because of the interesting topological aspects, most recent experimental and theoretical investigations have focused on the electronic band structure of specific surface orientations to identify the topological character \cite{jiang_observation_2013,xu_surface_2013,denlinger_temperature_2013, xu_direct_2014,kang_band_2013,kim_termination_2014}. 

\subsection{Typical surface characteristics in rare earth compounds}
Although some electronic characteristics of the surface electronic structure can be explained by the topology \cite{wolgast_low-temperature_2013}, there are the unique and well-known surface properties in rare earth compounds to be clarified \cite{oh_sxrespes_1984, duo_ce_1996, kim_surface_1997, duo_surface_1998, reinert_photo_1998, schmidt_xray_2005,suga_kondo_2005, ehm_high_2007, suga_unraveling_2009}. In the photoemission results, the typical surface peaks appear 1$\sim$2\,eV at higher binding energy \EB~than the bulk peaks; moreover, the valence of the rare earth ion $V_{RE}$ reveals a smaller value at the surface than in the bulk, \textit{e.g.} as in TmSe and several Ce- and Yb-compounds \cite{oh_sxrespes_1984, kim_surface_1997, reinert_photo_1998}. It indicates that the $f$ occupation \nf~of the surface region is larger than that of the bulk. This discrepancy appears not only at the top surface, but also in the subsurface region, whose existence has been proved by varying the probing depth in photoemission experiments \cite{reinert_photo_1998,suga_kondo_2005, ehm_high_2007, suga_unraveling_2009}. For instance, in YbInCu$_4$, the thickness of the subsurface region has been estimated to be 1\,$\sim$\,3 unit cells \cite{schmidt_xray_2005, Suga_potential_2008, suga_unraveling_2009}. Because of this unique discrepancy, the topological surface states in the rare-earth compounds can be differently related to the bulk states \cite{kim_termination_2014, seibel_connection_2015}. In general, $V_{RE}$ typically shows a time-dependent change after the cleavage toward higher valence, \textit{e.g.} $\textit{V}_{\rm RE_{surf}}$\,$\rightarrow$\,3 for TmSe and Yb- compounds. 

 The discrepancy in the valence between surface and bulk has been attributed to a reduced coordination, and the time dependence has been attributed to contamination \cite{braicovich_high_1997,duo_surface_1998,oh_sxrespes_1984}. However, \SB~shows almost opposite behavior. Below 50\,K the bulk Sm valence \vl~is $\sim$\,2.52  \cite{mizumaki_temperature_2009,yamaguchi_different_2013}, whereas the surface valence is higher than that of the bulk one \cite{Phelan_Correlation_2014}. In other words, \nf~of the surface is lower than the bulk one. This opposite tendency has been also detected in another mixed-valent Sm compound SmOs$_{4}$Sb$_{12}$ \cite{yamasaki_coexistence_2007}. 

\subsection{Terminations and time dependence in hexaborides}
In hexaborides, the (100) cleaved surface is a polar surface due to the alternating ionic charged planes consisting of the boron- and metal- terminations \cite{zhu_polarity_2013}. The charged planes producing the electrostatic potential have been suggested to explain a slow charge transfer in a scale of hours from one termination to the other. The time evolution in the valence band spectra of \SB~has been attributed to this electronic reconstruction. Thus, it is also important to investigate the electronic relaxation on the surface at different terminations in order to capture the essence of the surface electronic structure \cite{kim_termination_2014}. In particular, we study both Sm and B core-levels to clarify the charge redistribution near the surface region including the varying valence in the subsurface.

\subsection{Occupancy in the mixed valent insulator}
In this report, we characterize the time and temperature dependence of the spectroscopic features appearing on the cleaved (100) surface of \SB~\cite{jdenlinger_smb6_2014}. In particular, we focus on the Sm valence \vl, which reflects the $f$ occupation number (\nf\,$=$\,8\,$-$\vl). With \vl, we can additionally figure out the $d$ occupation because the sum of the $f$ and $d$ occupation has to be an even integer number for an insulator. Thus, \vl~is effective to investigate the charge redistribution. In the periodic Anderson model (PAM), \nf~is also a key parameter for the paramagnetic insulating phase, which reflects the gap size and the topological phase \cite{werner_interaction_2013}. Especially, we focus on the \vl~in the subsurface region. This region is structurally equivalent to the bulk and has the same number of nearest neighboring atoms, so the local parameters (e.g. the electron-electron coulomb interaction $U$, the single particle energy $\varepsilon_f$ of the occupied $f$\,states) should be almost identical to the bulk ones. Thus, the change in occupation for the subsurface region reflects the change in the other parameters, e.g. the hybridization strength, bandwidth of $d$\,states, \textit{etc}.

\section{\label{sec2}Method}
The single crystals were grown by the Al-flux method with samarium pieces (99.9 \%), boron powder (99.9 \%) and Al (99.999 \%). The mixture ratio of SmB$_6$ to Al was 1 to 50. For the growth, a vertical furnace was utilized, which was cooled down  from 1,500 $^{\circ}$C by 4.2 K/h. Sodium hydroxide (NaOH) solution was applied to remove the Al flux. The physical properties of the crystal have been demonstrated in the supplemental material of Ref.\,\cite{min_importance_2014}.
We performed x-ray photoelectron spectroscopy (XPS) experiments on \SB~using a VG Scienta R3000 electron analyzer and Al-$K_\alpha$ x-ray source ($h\nu$\,$=$\,1486.6\,eV) to study the time dependence of the Sm valence on the (100) surface. All single crystals for this investigations were glued on the holder and cleaved $in$ $situ$ by using a cleavage post. To ensure the temperature at the sample, we directly attached the holder on the manipulator. 
 
Moreover, in order to selectively probe the Sm ions on the different terminations, further core-level studies were carried out at two synchrotron radiation facilities. Soft x-ray photoelectron spectroscopy (PES) experiments were performed at the ASPHERE endstation of beamline P04 of PETRA III (Hamburg, Germany) equipped with a rotatable VG Scienta R4000 analyzer. This setup keeps a constant beam spot size of few hundred $\mu$m on the sample surface since the incident angle is fixed. The sample temperature was $T$\,$=$\,30\,K. The real-space XY mapping with core-levels was carried out for the entire cleaved surface to get reliable positions for each terminations during take-off angle measurements. 
Hard x-ray PES experiment were performed at the I09 beamline of the Diamond Light Source (Didcot, United Kingdom) where the endstations are equipped with a VG Scienta EW4000. All take-off angle measurements were performed at $T$\,$=$\,150\,K with the angular mode of the analyzer, which covers $\sim$\,60$^{\circ}$. In this setup, the same beam position and spot size on the sample surface are guaranteed during the measurement. The base pressures was always better than 3.0$\times$10$^{-10}$\,mbar.

\section{\label{sec3}Time dependence}
\begin{figure}
	\centering
		\includegraphics[width=0.5\textwidth]{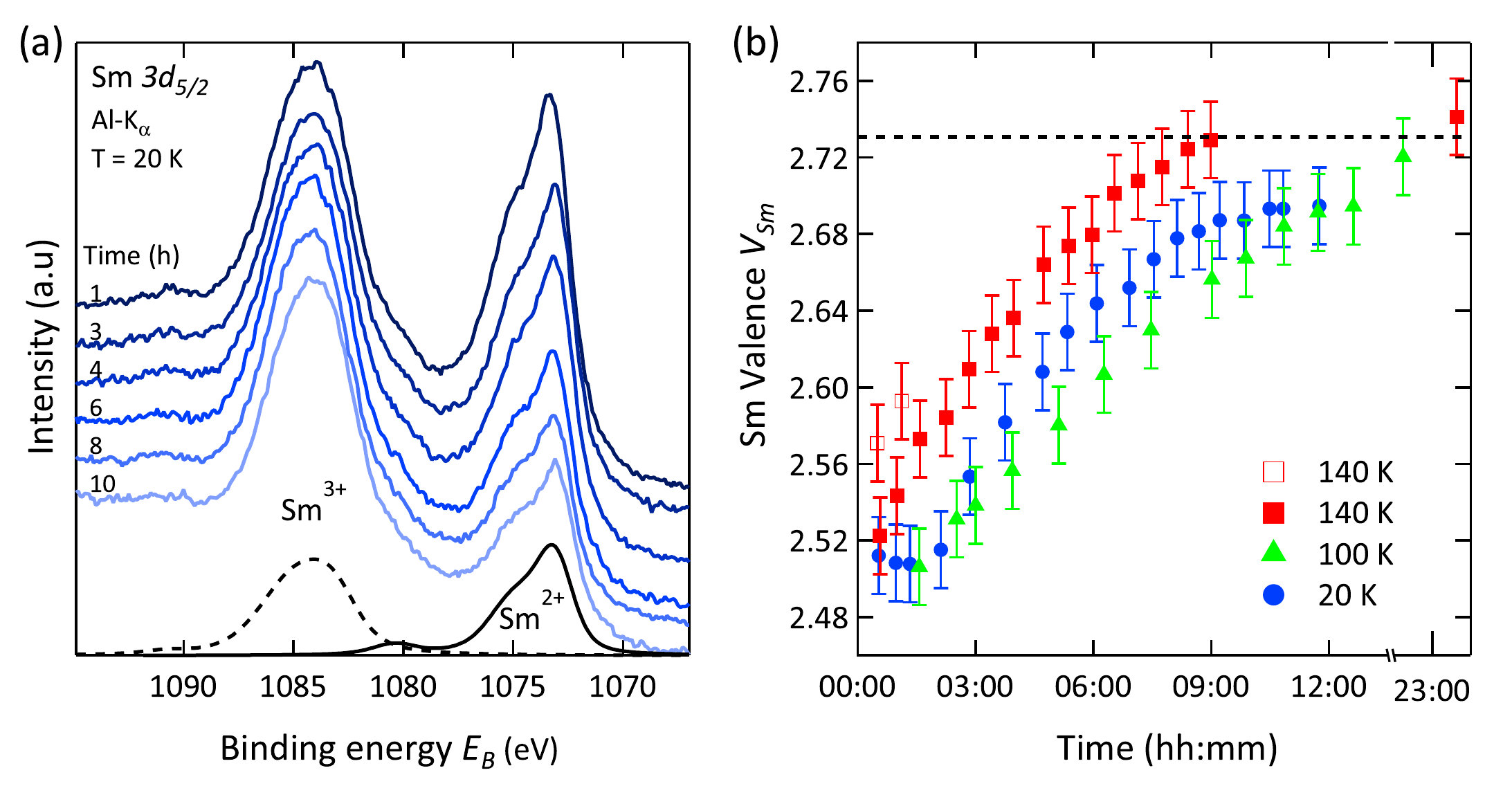}
			\caption{Time dependence of Sm valence $V_{Sm}$ after the cleavage. \SB~crystals were cleaved and measured at the same temperatures. (a) As an example of typical Sm $3d_{5/2}$ core-level spectra of \SB, we show the spectra of the sample cleaved and measured at $T$\,$=$\,20~K. They show two main contributions of \sth~and \stw since it is a homogeneous mixed valent system. The intensity of the \stw~peak relative to that of the \sth~reduces with time. The spectra at the bottom with solid and dashed lines represent the theoretical and instrumentally broadened spectral weights of the \stw~and \sth~contribution, respectively \cite{cho_surface_1999}. (b) $V_{Sm}$ \textit{vs.} time of four \SB~crystals that are obtained at the temperatures of 20~K, 100~K, and 140~K. The initial \vl~shows different values for different sample, such as the two different values at 140\,K (red squares), but the obtained valences are always less than 2.6. In the first $\sim$\,9\,h, it gradually increases, and saturates to $\sim$2.73 indicated by the black dotted line.}
	\label{fig1}
\end{figure}

In order to study the change in \vl~as a function of time after the cleavage, we have measured Sm $3d_{5/2}$ core-levels at a constant temperature by use of Al-$K_{\alpha}$ source. Each sample was measured at the same temperature as it was cleaved. During the measurement all external parameters were kept constant in order to study solely the time dependence in \vl. Due to the time evolution occurring in several hours \cite{zhu_polarity_2013}, this study requires a data acquisition within 15 min for each core-levels. The energy resolution with the chosen setup of the spectrometer was $\sim$1.5~eV. 

Fig.\ref{fig1}\,(a) shows the 3$d_{5/2}$ spectra of a sample cleaved and measured at 20\,K. \sth~and \stw~peaks appear at  \EB~of 1084\,eV  and 1073\,eV, respectively. The spectra are shifted vertically along the y-axis. With respect to \sth, the spectral weight for \stw~reduces dramatically within 9\,h. The spectral weight of \sth~and \stw~were determined by fitting with the theoretical line shapes of 3$d_{5/2}$ for both \stw~and \sth~\cite{cho_surface_1999,cho_electronic_2007,yamasaki_coexistence_2007} and a Shirley background \cite{hufner_photoelectron_2003}. By using the theoretical \stw~and \sth~spectra as shown at the bottom of the graph with black lines, the overlapped contributions of \stw~and \sth~can be separated reliably (\EB\,$=$\,1080\,eV).

We reproduced the results for additional samples and temperatures (Fig.\ref{fig1}\,(b)). At each temperature, the spectra show the lowest \vl~directly after the cleavage. As time goes on, \vl~increases, and saturates at 2.73 after $\sim$\,9\,h. Note that the initial \vl~differs for every sample even when the crystals got cleaved at the same temperature (see the red filled and unfilled squares). There is a big rise in \vl~between 3 and 9 h for all cases although they were cleaved and measured at different temperatures and base pressures. 

In fact, after the first 12\,h under UHV conditions, the samples were exposed to a pressure burst ($p > 1\times10^{-6}$\,mbar) due to the natural vaporization of residual gases by warming up the cryostat to room temperature. After another 12\,h, we cooled the samples to their initial temperatures, and measured \vl~again. Surprisingly, the Sm ions show almost the same \vl~as the saturated values before the pressure change. This indicates that the effects due to contamination from residual gases are negligible in the valence change.

Since the bulk \vl~is constant at a fixed temperature \cite{mizumaki_temperature_2009}, the major changes seen in Fig.\ref{fig1} are probably due to the Sm ions near the surface region. Unlike Yb compounds \cite{reinert_photo_1998,suga_kondo_2005, ehm_high_2007, suga_unraveling_2009}, the initial \vl~of \SB~shows similar or sometimes higher values than the bulk \vl~(see ref.\cite{mizumaki_temperature_2009}). Moreover, \vl~rises with time, and it saturates to a non-integer value. 

\section{\label{sec5}Spectral weight analyses for the adsorption layer}

\begin{figure}
	\centering
		\includegraphics[width=0.5\textwidth]{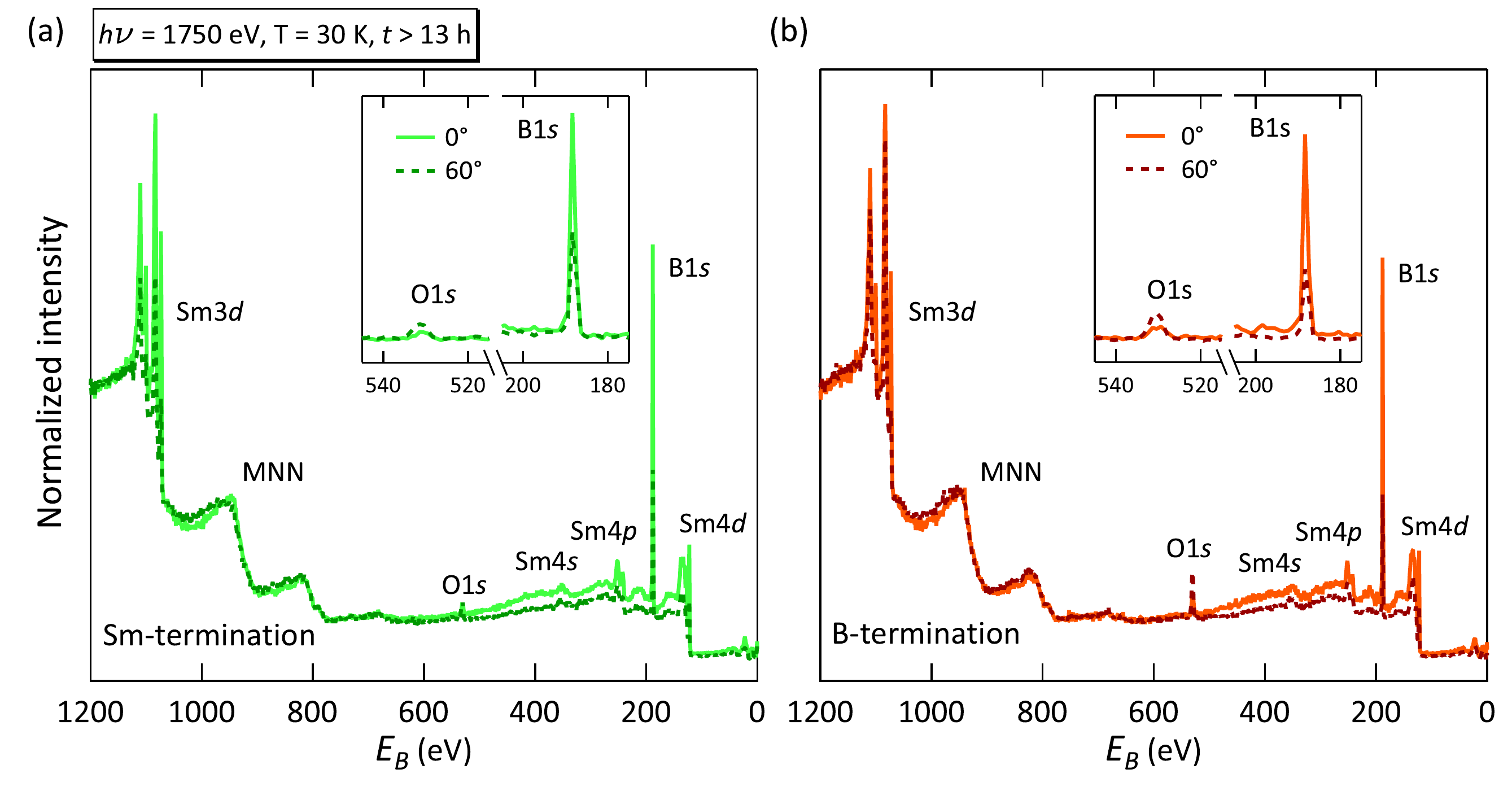}
			\caption{Overview spectra of all reachable core levels including the oxygen O1$s$ peak measured at PETRA III synchrotron radiation facility ($h\nu$\,$=$\,1750\,eV, $T$\,$=$\,30\,K). (a) and (b) show the change of peak intensities for the two corresponding terminations, Sm and B, after oxygen contamination of the sample under UHV condition for 13\,h (solid line, $\vartheta$ = 0$^\circ$) and for 26\,h (dotted line, $\vartheta$ = 60$^\circ$). During all these measurements no notable C 1$s$ intensity was observed. The insets emphasize the reduction in intensity of the B 1$s$ and the enhancement of the O1$s$ peak after time. Considering the intensity ratio I$_{\rm O}$($\vartheta$, t)/I$_{\rm B}$($\vartheta$, t) of each termination (Sm or B), the oxygen layer thickness is determined.}
	\label{fig4}
\end{figure}

The time evolution happens in a period of several hours. Thus, the implication of adsorbed residual gas on the sample as well as that of oxidation has to be investigated. Additional adsorbate layers on the surface reduce the probing depth from the cleaved surface, which might also reduce the bulk signal of \SB. In particular, oxidation can give rise to higher \vl\,$\rightarrow$\,3 since the common chemical phase for Sm oxides is Sm$_2$O$_3$. 

Fig.\,\ref{fig4} depicts all the relevant core-levels of \SB~for the two different terminations (See also Sec\,\ref{sec4}), which were taken 13\,h after the cleavage. No C\,1$s$ has been detected in the sample area, so the contamination due to the carbon based molecules and intrinsic C-doping is excluded \cite{Phelan_Correlation_2014}. The spectra were taken with a photon energy of 1750\,eV, since in this energy range the O 1$s$ core-level is not superimposed by MNN Auger peaks. Here, the take-off angle $\vartheta$, which is the angle between the detector and the surface normal, has been varied from $\vartheta$\,=\,0$^\circ$ to 60$^\circ$. The insets show the O\,1$s$ and B\,1$s$ peaks, which are important to estimate the oxygen layer thickness.

A simple model was considered to determine the growth of an oxygen layer on the substrate \cite{briggs1983practical}: the intensities of an outgoing electron from the oxygen layer $\mathrm{I}_{O}$ and from the actual substrate (\SB) $\mathrm{I}_{B}$ are given by 

\begin{equation}\label{Oint}
\mathrm{I}_{O}=\alpha_{O}I^{0}_{xray}(1-e^{-d/\lambda_{O}\cos\vartheta}),\hspace{2ex}
\mathrm{I}_{B}=\alpha_{B}I^{0}_{xray} e^{-d/\lambda_{B} \cos\vartheta}
\end{equation}

\bigskip

where $\alpha_{O}$, $\alpha_{B}$ are the atomic sensitivity factors of the O\,1$s$ and B\,1$s$ orbitals respectively \cite{moulder1992handbook}. The IMFPs of the photoelectrons from O\,1$s$ and B\,1$s$, $\lambda_{O}$ and $\lambda_{B}$, depend on their different kinetic energies.

From Eq.\eqref{Oint} we get the intensity ratio

\begin{equation} \label{ratio}
\dfrac{\mathrm{I}_{O}}{\mathrm{I}_{B}}=\dfrac{\alpha_{O}}{\alpha_{B}}\dfrac{(1-e^{-d/\lambda_{O}\cos\vartheta})}{e^{-d/\lambda_{B} \cos\vartheta}}
\end{equation}
The atomic sensitivity factors of the 1s orbitals are given by $\alpha_{O}=0.711$ and $\alpha_{B}=0.159$ \cite{moulder1992handbook}, and the IMFP, $\lambda_{O}=22.37$\,\AA~ and $\lambda_{B}=27.5$\,\AA, were taken from the NIST Standard Reference Database \cite{powell2000nist}.
Since $(\lambda_{O}-\lambda_{B})/(\lambda_{O}\,\lambda_{B})$\,$<<$\,1, the following approximation is applicable:
\begin{equation} \label{appr}
e^{(-d/\cos\vartheta)(\lambda_{O}-\lambda_{B}) / (\lambda_{O}\lambda_{B})} \approx 1
\end{equation}

and we get a rather simple formula for the oxygen layer thickness
\begin{equation} \label{thick}
d=ln(\dfrac{\mathrm{I}_{O}}{\mathrm{I}_{B}}\dfrac{\alpha_{B}}{\alpha_{O}}+1)\lambda_{B} \cos\vartheta.
\end{equation}

\bigskip

To determine the intensities $\mathrm{I}_{O}$ and $\mathrm{I}_{B}$, the spectra in Fig.\,\ref{fig3} were Shirley background corrected and the peak areas integrated. This results in the layer thicknesses given in Table 1:

\begin{center}
    \begin{tabular}{| l |c |c |}
    \hline  \label{Olayers} && \\ 
     & Sm Termination & B Termination \\ \hline
    $d$ [\AA] after 13\,h & 0.51 & 0.82 \\ \hline
    $d$ [\AA] after 26\,h & 1.07 & 2.46 \\ \hline
    \end{tabular}
\end{center}

This means that the oxygen layers are more than one order of magnitude thinner than the IMFP, and the sample surface has not yet been covered with a monolayer of the oxide, especially before the saturation regime of Fig.\,\ref{fig1}. Thus, the adsorption of residual gases on the surface cannot explain the change in the valence.

\section{\label{sec4}Spatial dependence}

\subsection{Sm and B terminations and surface Peaks}
\begin{figure}
	\centering
		\includegraphics[width=0.5\textwidth]{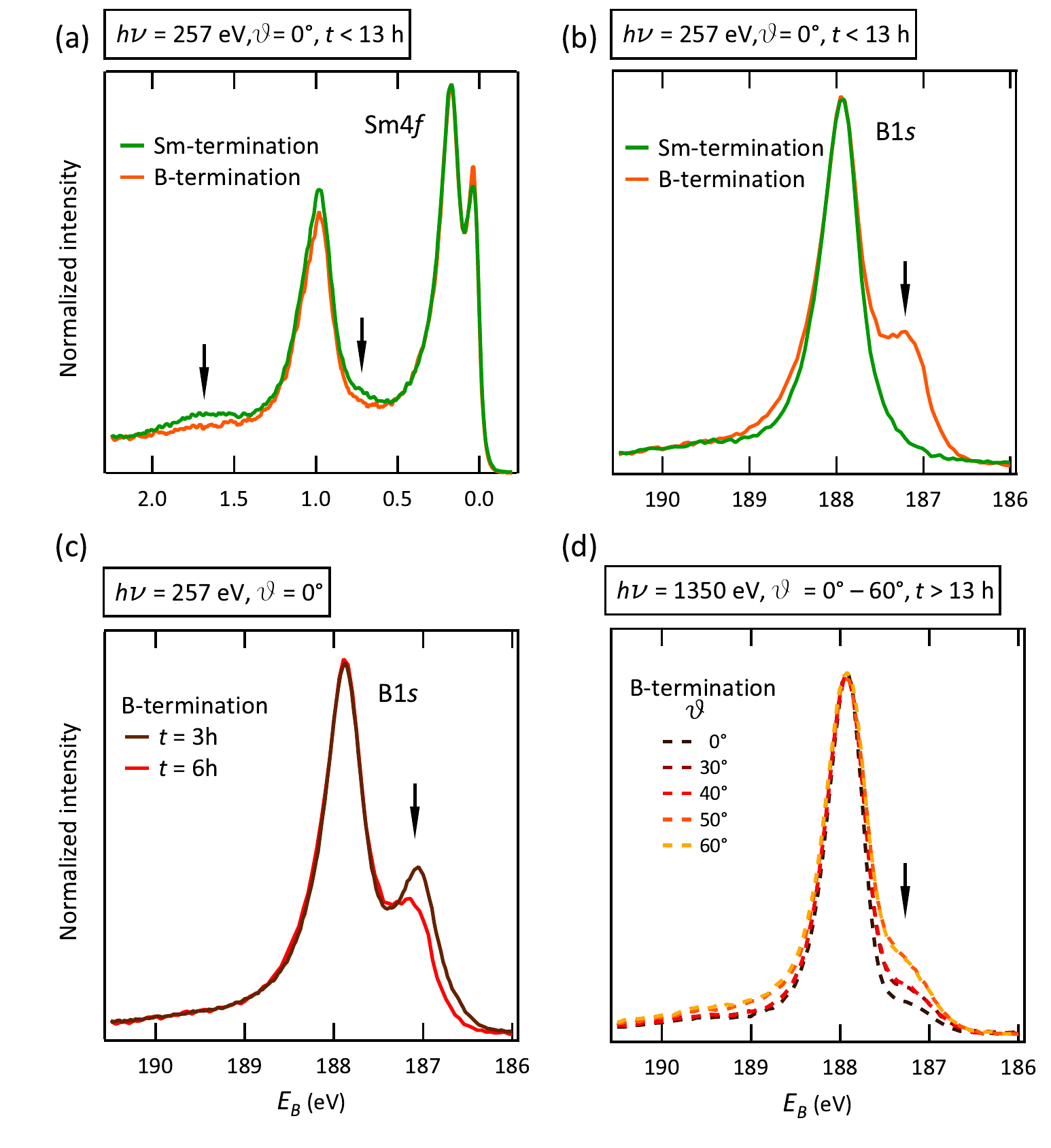}
			\caption{Signatures of the Sm- and B- terminated surfaces appearing in (a) \stw\,4$f$ and (b) B\,1$s$ spectra. The spectra were taken with $h\nu$ = 257\,eV and selected from a real-space map of the cleaved surface (See also Fig.\,\ref{fig3} (a)-(b)). Sm- and B- terminated surfaces show broad peaks at higher and lower \EB~of $\sim$\,1\,eV from the main peaks, respectively (black arrows). (c) Aging of the B\,1$s$ surface peak. Within several hours, the surface peak decreases and saturates at a certain height. (d) Angle dependence in the B\,1$s$ measured with $h\nu$ = 1350\,eV at the B-terminated area. At higher angles, the low \EB~shoulder in the B peak get stronger, which confirms the surface origin of the shoulders.}
	\label{fig2}
\end{figure}

Because \vl~varies remarkably with time, Sm ions can be involved in the electronic reconstruction between Sm- and B-terminations \cite{zhu_polarity_2013}. We investigated the spatial dependence on the cleaved surface. First, the entire surface was monitored with \stw\,4$f$ and B\,1$s$ spectra using the photon energy of 257\,eV. The sample position has been varied below the small beam spot, which is available at the synchrotron radiation facility PETRA III (see Sec.\,\ref{sec2}). Fig.\ref{fig2}\,(a) and (b) represent the typical line shape of \stw\,4$f$ and B\,1$s$ spectra, respectively. Besides the sharp main peaks \cite{chazalviel_study_1976,denlinger_advances_2000,min_two_2015}, there are additional broader peaks (black arrows) for each core-level. These broad contributions have been assigned to the surface spectral features \cite{jdenlinger_smb6_2014, heming_surface_2014,patil_unusual_2011} (see also Fig.\ref{fig2}\,(d)), which are utilized to assign the Sm- or B-terminated regions. The intensity of these surface peaks decrease with time as shown in Fig.\ref{fig2}\,(c).

If we define a probing depth to be the product of the inelastic mean free path (IMFP) and cosine of the take-off angle, $\lambda\cdot\cos\vartheta$, the probing depth for $h\nu$ = 1350\,eV and $\vartheta$ = 60$^{\circ}$ will be 10.8\AA, whereas that for $h\nu$ = 257\,eV and $\vartheta$ = 0$^{\circ}$ is 4.8 \AA~\cite{powell2000nist}. Thus, the contribution of the shoulder could be reduced by more than 50\%~compared to that of (c). Our take-off angle study on the boron peak confirms the surface origin of the shoulder (Fig.\,\ref{fig2}\,(d)) because the spectra at the higher angle, which are more surface-sensitive, shows a higher shoulder. 

\subsection{\label{sec4.2}\vl~at the Sm and B terminations}
\begin{figure}
	\centering
		\includegraphics[width=0.5\textwidth]{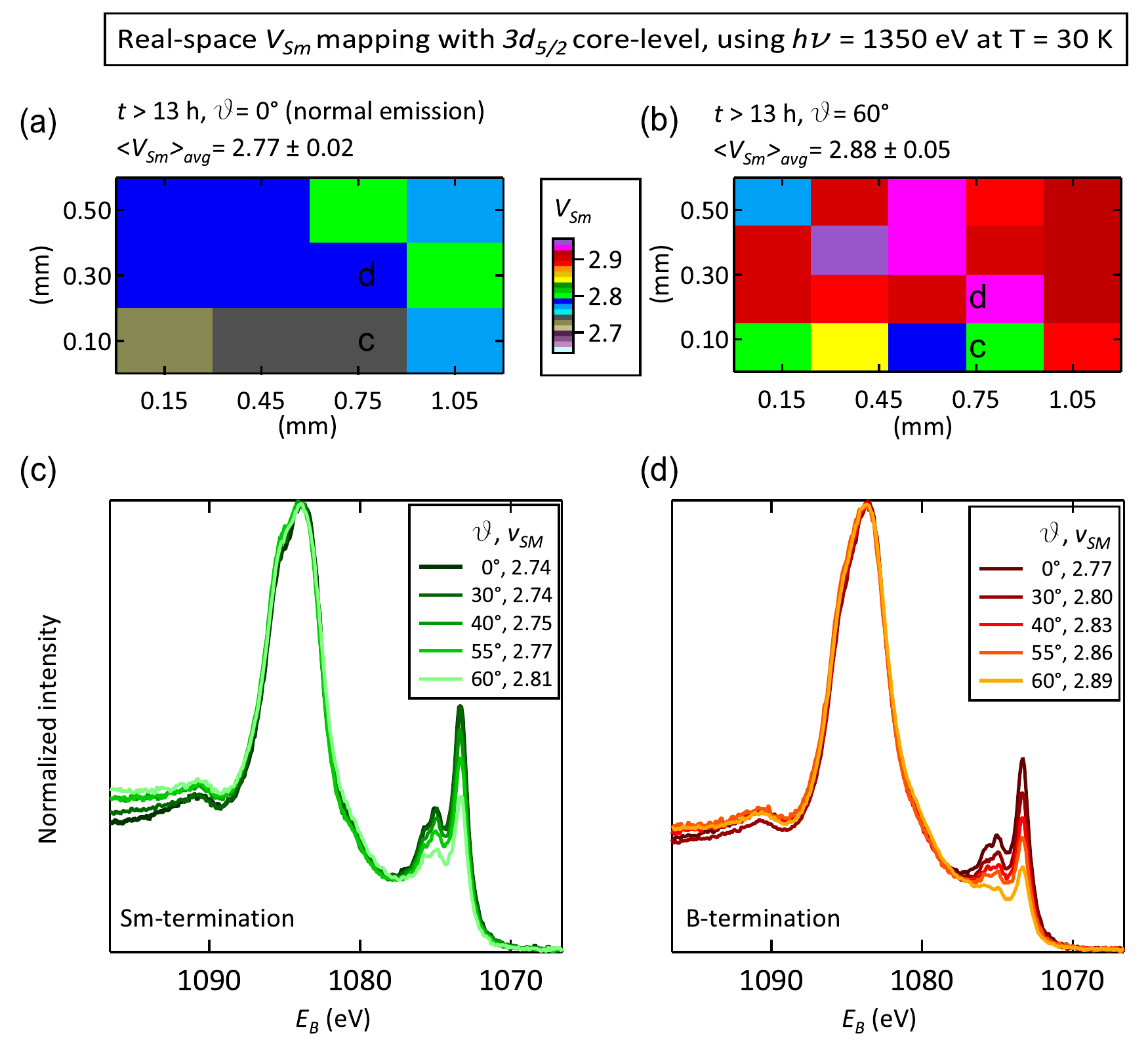}
			\caption{(a-d) Using the $\mu$m size of the beam-spot from the synchrotron radiation facility, the spatial dependence of \vl~was investigated on the (100) surface with the photon energy of 1350~eV at $T$\,$=$\,30\,K , which has been cleaved in the UHV 13 h before. The area of 1.2$\times$0.6 mm$^2$ of the cleaved surface was scanned. (a) The \vl~map was obtained in the normal emission. The average \vl~of this area is 2.77, which is similar to the results in Fig.\ref{fig1}. However, small spatial variation has been detected throughout the surface.  (b) The valence map was obtained at the $\vartheta$ = 60$^{\circ}$ off normal. The average valence shows a higher value (\vl\,$=$\,2.88) than that in (a), which indicates the Sm ions on the surface have a higher valence. (c) - (d) The angle dependence in \vl~taken at the two different terminations, whose locations are indicated by the labels \textbf{c} and \textbf{d} in the map (a) and (b).}
	\label{fig3}
\end{figure}

After the experiments with the surface-sensitive excitation energy at 257\,eV (Fig.\,\ref{fig2}), we have changed the photon energy to $h\nu$\,$=$\,1350\,eV in order to investigate the spatial dependence of \vl~on the Sm\,3$d_{5/2}$ core-levels. When the sample surface was mapped with this photon energy, it had been already exposed to the UHV longer than 13\,h. Thus, the sample should be in the saturated \vl~regime as indicated with the dotted line in Fig.\,\ref{fig1}\,(b). 

Fig.\,\ref{fig3}\,(a) shows the map of the \vl~estimated from the Sm\,3$d_{5/2}$ at normal emission. As shown in the color-scale placed in the middle of Fig.\,\ref{fig3}, different spots show slightly different \vl. The average of \vl~for the surface is 2.77, which is similar to the saturated value in Fig.\,\ref{fig1}\,(a). Photoelectrons from \stw\,3$d$ have an IMFP of 7.86 \AA~\cite{powell2000nist}. By moving the analyzer 60$^\circ$ off normal emission, we can reduce the probing depth by a factor of two and obtain the more surface-sensitive \vl~maps (Fig.\,\ref{fig3}\,(b)). The average \vl~is higher (\vl\,$=$\,2.88) than that at normal emission. Thus, the surface region shows a significant increase in \vl~within 4 \AA~depth from the surface.

The positions marked \textbf{c} and \textbf{d} in Fig.\,\ref{fig3}\,(a-b) correspond to Sm- and B-terminations, respectively, which are the same positions where the spectra in Fig.\,\ref{fig2}\,(a-b) were obtained. We have performed angle-dependent measurements of the 3$d_{5/2}$ core-level on this two particular spots (Fig.\,\ref{fig3}\,(c-d)), respectively. The Sm-terminated region shows a lower \vl~value than the B-terminated region by 0.05. Both terminations show higher \vl~near the surface region than in the bulk. The difference of the valence in the two terminations remains similar for all angles. Moreover, the background intensity and the peak shapes depend on the terminations, but the whole area show a mixed valency.

\section{\label{sec6}Subsurface region}
\begin{figure}
	\centering
		\includegraphics[width=0.5\textwidth]{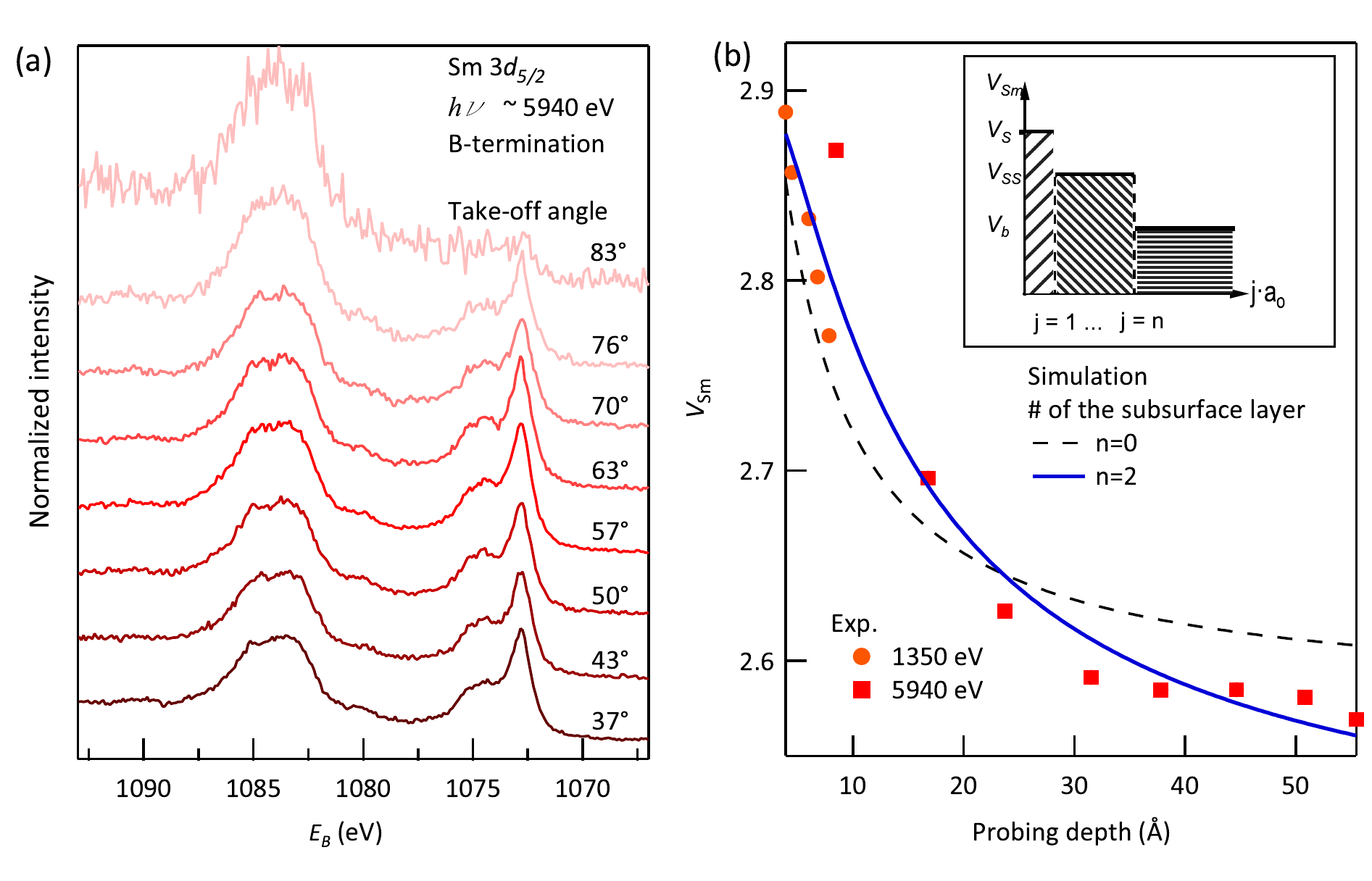}
			\caption{(a) Take-off angle measurement at Sm 3$d_{5/2}$ using the angular mode of the analyzer. The sample was cleaved more than 9\,h before the measurement, and its temperature was kept to 150\,K. The photoelectron from the 3$d$ states has a IMFP of $\lambda$\,$=$\,69.45 \AA. The spectral weight ratio of \stw~to \sth~peaks decreases with increasing angle $\vartheta$. (b) The \vl~values from (a) are depicted with the red square markers as a function of the probing depth, $\lambda$$\cdot$$\cos$$\vartheta$. In addition to the HAXPES results, the values obtained in Fig\,\ref{fig3}\,(d) are shown with the orange circles. The inset in (b) shows the model that we use to estimate the subsurface thickness. Based on the our results and literature \cite{mizumaki_temperature_2009}, we have following constraints. The bulk \vl ($V_{b}$) is around 2.5, but certainly below 2.6, and the surface and subsurface \vl ($V_{s}$ and $V_{ss}$) are around 2.89 but less than 3. If the subsurface does not exist, our model result strongly deviates from the experimental values as shown with the dashed line although the $V_{s}$\,$=$\,3. Our best results are achieved when there are two unit cell of the subsurface region beneath the surface (blue line).  
}
	\label{fig5}
\end{figure}

In order to systematically study \vl~as a function of depth, we performed take-off angle measurements at the Sm 3$d_{5/2}$ core-level on the B-terminated surface. Due to the unique analyzer setup of the endstation at the I09 beamline at DIAMOND, spectra for different take-off angles $\vartheta$ could be measured simultaneously. The sample temperature was 150\,K and the photon energy was 5940\,eV. The spectra in Fig.\,\ref{fig5}\,(a) were taken more than 9\,h after the cleavage ($t$\,$>$\,9\,h). The spectral weight ratio of \stw~to \sth~decreases with increasing $\vartheta$; this indicates that \vl~increases towards the surface, which is consistent with the results from Fig.\,\ref{fig3}.

Since HAXPES can be very bulk-sensitive, the spectra taken at $\vartheta$\,$<$\,60$^\circ$ show only small changes. Regarding the kinetic energy of 4.8\,keV for the 3$d$ photoelectrons, the theoretical IMFP is $\lambda$\,$=$\,69.45 \AA~\cite{shinotsuka_calculation_2015}. Since the probing depth equals to $\lambda$$\cdot$$\cos$$\vartheta$, we can plot \vl~as a function of depth as shown with the red squares representing the HAXPES data in Fig.\,\ref{fig5}\,(b). In order to monitor the full-range variations of \vl~in depth, the \vl~values in Fig.\,\ref{fig3}\,(c) (orange-circle) are also plotted with their corresponding IMFP, $\lambda$\,$=$\,7.86 \AA~($E_K$\,$\approx$\,280\,eV). Although the spectra in Fig.\,\ref{fig3}\,(c) were taken at $T$\,$=$\,30\,K, unlike that at 150\,K for Fig.\,\ref{fig5}, we can put both \vl~data together in one graph because \vl~near the surface region is insensitive to the temperature after 9\,h.

To estimate the thickness of the subsurface region, which shows different \vl~from the bulk and the surface, we construct a model as shown in the inset of Fig\,\ref{fig5}\,(b), which has been applied to the other rare earth compound \cite{suga_unraveling_2009}. Three different regions are considered: one for the top surface, $V_s$, another for the subsurface region with the number of unit cells $n$, $V_{ss}$, and the other is for the bulk, $V_b$. The obvious limits are $V_s$\,$<$\,3, $V_{ss}$\,$<$\,3, and $V_b$\,$<$\,2.6 to be consistent with the experimental results \cite{mizumaki_temperature_2009}. We set the lattice constant $a_o$ of \SB~to be 4.13 \AA, and describe in term of the unit cell because there is only one Sm ion per unit cell.

The average valence that we get from the Sm core-level measurement can be estimated from this model. Since each unit cell contributes a different weight to a spectrum, we include the IMFP in the attenuation: 
\begin{equation} \label{atten}
\text{Attenuation}\propto e^{-jA},\hspace{1ex}\text{where } A \equiv \dfrac{a_o}{\lambda_{Sm}\cos\vartheta} \text{, \& } j = \text{integer}
\end{equation}
$a_o$ is the lattice constant, $\vartheta$ is the take-off angle from the normal emission, and $j$ is the unit cell index. Thus, the $j\cdot\,a_o$/cos$\vartheta$ will be the travel length of the photoelectron from the $j$-th unit cell. $\lambda_{Sm}$ is the IMFP of the photoelectron from the 3$d_{5/2}$ state. We have used the density of \SB~(5.07 g$/$cm$^{3}$) to get the theoretical IMFP. The IMFP of \stw~and \sth~are presumed to be equal. Based on the parameters and factors, the average \vl~can be obtained from the following equation:
\begin{equation} \label{subsurf}
<V_{Sm}>_{exp} =\frac{V_{s} + \sum^{n}_{j=1}V_{ss} e^{-jA} + \sum^{\infty}_{j=n+1}V_{b} e^{-jA}}{\sum^{\infty}_{j=0}e^{-jA}} 
\end{equation}
The denominator is the total spectral weight.

This equation is simplified as follows:
\begin{equation} \label{subsurf1}
<V_{Sm}>_{exp} = V_{s} (1-e^{-A}) + V_{ss} (1-e^{-(n+1)A}) + V_{b} e^{-(n+1)A}
\end{equation}

We fit the experimental data in Fig.\,\ref{fig5}\,(b) with the constraints given above and the various integer values for $n$. For $n$\,$=$\,0, the best result (dashed line) is obtained when $V_s$ is set to the highest possible valence of 3, which objects our results for the measurements with $h\nu$\,$=$1350\,eV. Nevertheless, the fitting shows no satisfying result. The results for $n$\,$=$\,1 shows a little bit worse result than that of $n$\,$=$\,2 without violating the constraints. However, the resulting $V_{ss}$ corresponds to 3, which violating our observation as well. The least-squares fitting shows the best result for $n$\,=\,2, which gives $V_{s}$\,$=$\,2.906\,$\pm$\,0.086, $V_{ss}$\,$=$\,2.872\,$\pm$\,0.126, and  $V_{b}$\,$=$\,2.480\,$\pm$\,0.027. Our $n$ value is consistent with the theoretical prediction in Ref.\cite{kim_termination_2014}. For $n$\,$>$\,2, $V_b$ becomes lower than 2.47, which starts to strongly deviate from bulk-sensitive results \cite{mizumaki_temperature_2009}. 

\section{Discussion}
We have evaluated Sm valences with two different methods in order to double-check our estimated values. The first uses the theoretical multiplet structure, and the other is based on a simple Shirley background correction. All the Figures in this study show the results obtained from this first method. With the theoretical multiplet structure of 3$d_{5/2}$ \cite{cho_surface_1999,cho_electronic_2007,yamasaki_coexistence_2007}, we can obtain each line shapes for \stw~and \sth~after small corrections in peak positions and widths. The corrections are made to fit our HAXPES spectra with theoretical ones considering the Shirley background. This method usually gives the most reliable values, but the spectra in Fig.\,\ref{fig3}\,(d) are difficult to fit with only three components within the multiplet structure. For instance, \stw~showing significant peak broadening with time and strong intensity of the background, there can be some leftover spectral weights with the three component analysis.
Besides the fitting method, we have analyzed it by simply subtracting the Shirley background and strictly separate \stw~and \sth~by setting their border at 5\,eV in high \EB~from the main peak of \stw. This method does not consider the overlapped region of \stw~and \sth at \EB\,$=$\,1080\,eV, which produces $\sim$ 7.5\,$\%$ error in estimating \stw~spectral weight. However, the full spectral weight of $3d_{5/2}$ was considered for the estimations. From both analyses, we have observed consistent time evolution like in Fig.\,\ref{fig1}\,(b) although two methods give different \vl~values by less than 0.05.

A drastic change in \vl~occurs within 9\,h after the cleavage and saturates at a value of \vl~= 2.73. We can exclude a chemical reaction with residual gases by the following reasons. First, from our results of the O\,1$s$ and B\,1$s$ spectra obtained from the synchrotron radiation facilities, we analyze the thickness of the oxygen layer on the surface. However, the layer thickness of oxygen is negligible within 13\,h after cleavage. Moreover, we performed the same experiment at different temperatures so that also the base pressure for each measurement varied. 
If contamination affects \vl, there should be a different behavior for the individual measurements. However, \vl~reaches to the saturated value in the similar time-scale. It indicates \vl~change is insensitive to the base pressure. Furthermore, we performed some of the measurements at low temperatures $T$\,$\sim$\,20\,K. Thus, processes including chemical bonding of the Sm ion with the adsorbate are hindered since the temperature is not high enough to overcome the activation energy. 

Instead, we see a close connection between the changes in the valence \vl~and the decrease of the B and Sm surface peaks with time (Sec.\,\ref{sec4}). Both effects happen on the same timescale, which also has been observed in the valence band study \cite{zhu_polarity_2013}. Since we have confirmed that \vl~near the surface is higher than for the bulk, we can naturally model the distribution of the \vl~as a function of depth. We tried several models, \textit{e.g.} with a linear variation in \vl~from surface to bulk, but the best result is achieved with the current model using a subsurface region. Although we used a simplified model, we get very good agreement to our experimental results as well as to the theoretical prediction. We determined the thickness of the subsurface region to be two times of the lattice constants $a_o$ of SmB$_6$. 
Note that \vl~varies slightly with the location on the surface, but significantly with the depth, \textit{e.g.} $V_s$ \textit{vs.} $V_b$. It indicates that the subsequent charge redistribution occurs between the surface and subsurface region in this very slow timescale together with the reduction in the surface shoulders in the core levels. Therefore, our result emphasizes that, in order to understand the discontinuity of the polar (100) surface of \SB, the instability of the Sm valence, $i.e.$ the $f$ occupancy, should be included in theoretical considerations, as already demonstrated in Ref.\cite{kim_termination_2014}.

\section{Summary}
We have studied the change of \vl~in \SB~on (100) cleaved surface with respect to different parameters such as cleavage quality, time, temperature, probing depth and surface terminations. The time-dependent study of \vl~indicates that a strong and slow charge redistribution occurring near the surface, which originates from neither adsorption nor reaction with residual gases. Moreover, our study has revealed the following characteristics of \vl~at the two terminations (B and Sm): (1) both terminations show apparent higher \vl~on the surface than in the bulk. (2) \vl~at the B-terminated area is slightly higher than that at the Sm-termination. Finally, we have clarified the existence of a subsurface region by estimating its extension with a model that gives a thickness of about two unit cells. Our results indicate that the main charge redistribution occurs from the top surface to the subsurface region, which induces higher \vl~in the subsurface region. Our findings confirm the importance of the $f$ states in the electronic reconstruction in the mixed valent \SB.  

\section{Acknowledgments}
This research was supported by the DFG (through SFB 1170 "ToCoTronics", projects C06 and the project RE 1469/8-1). B.K.C. and B.Y.K. were supported by National Research Foundation of Korea (NRF) grants funded by the Korean government (MSIP; Grants No. 2011-0028736 and Bank for Quantum Electronic Materials-BQEM00001). Parts of this research were carried out at the light source PETRAIII at DESY, a member of the Helmholtz Association (HGF). We would like to thank M. Kall\"ane, A. Quer, E. Kr\"oger, Jens Viefhaus for assistance in using beamline P04. We also thank Diamond Light Source for access to beamline I09 (SI11952) and especially T.-L. Lee and C. Schlueter for experimental and technical support on site. C.H.M. is grateful for the helpful discussion with S. Suga.

\bibliography{PhMag}

\end{document}